# Unanticipated synthesis of hexagonal GeSn alloys using pressure and temperature

George Serghiou,*[a] Hans Josef Reichmann,[b] Gang Ji,[c] Laurence Nigay,[d] Jonathan P. Wright,[e] Daniel J. Frost[f] and Gus Calder[g]

[a] Dr. G. Serghiou
School of Engineering
University of Edinburgh
Kings Buildings, Robert Stevenson Road EH9 3FB Scotland UK
E-mail: george.serghiou@ed.ac.uk
[b] Dr. H. J. Reichmann
Deutsches GeoforschungsZentrum, GFZ
Telegrafenberg,14473 Potsdam Germany
[c] Dr. G. Ji
Univ. Lille, CNRS, INRA, ENSCL, UMR CNRS 8207
UMET, Unité Matériaux et Transformations
59000 Lille France
[d] Prof. Dr. L. Nigay
University Grenoble Alpes - UGA
Laboratoire d'Informatique de Grenoble
38401 Saint Martin d'Hères France
[e] Dr. J. P. Wright
The European Synchrotron, ESRF
71 avenue des Martyrs, 38000 Grenoble France
[f] Prof. Dr. D. J. Frost
Bayerishes Geoinstitut, University of Bayreuth
95540 Bayreuth Germany
[g] G. Calder
School of Geosciences
University of Edinburgh
Kings Buildings, West Mains Road EH9 3JW Scotland UK

**Abstract:** Despite their electronic dominance, cubic diamond Si and Ge, are optoelectronically deficient. Recent work indicates however that a volume-expanded hexagonal Ge modification can exhibit intensely sought, superior optoelectronic characteristics. If larger Sn could form a hexagonal solid solution with Ge this would achieve this expansion. However, this was not anticipated because Ge and Sn are unreactive at ambient conditions, Sn does not have an ambient hexagonal symmetry and only cubic or tetragonal binary modifications could be prepared under any conditions. This state of affairs is categorically changed here by subjecting Ge and Sn to pressures of 9 and 10 GPa and temperatures up to 1500 K using large-volume press methods. Synchrotron angle-dispersive X-ray diffraction, precession electron diffraction and chemical analysis using electron microscopy, reveal ambient pressure recovery of hexagonal 2H, 4H and 6H Ge-Sn solid solutions ($P6_3/mmc$). Formation of this new binary materials landscape is correlated with Sn uptake, with the hexagonal symmetry being accessible below 21 at% Sn and the cubic diamond symmetry above this value. The findings create fertile routes to advanced materials, both in producing needed crystal symmetries based on composition and opportunity to tune properties based on crystal symmetry, composition and stacking sequence for optoelectronic applications.

Cubic diamond structured Si is the principal material component of the information processing units controlling virtually all electronic devices. The ever-growing demand for larger-scale, faster, less energy consuming, more varied information processing capability and connectivity and intrinsic light emitting applications is a strong driver for extending Si's microelectronic dominance to optoelectronics.[1,2] However, neither cubic diamond Si nor its established electronics industry cubic diamond complements, Ge or SiGe, are efficient absorbers or emitters of light because of their indirect band-gaps.[3,4,5,6] Thus the prime target is to develop materials which do have a direct band-gap, are efficient light emitters, are tunable and are compatible with group IVA cubic diamond (Si,Ge) to ensure efficient coupling of light transmission with electrical transmission.

The challenge hence is to create new materials within the same group IVA, exhibiting fundamental and tunable direct band gaps. One high potential structural symmetry is hexagonal. This is because hexagonal Ge ($P6_3/mmc$), unlike cubic diamond Ge ($Fd\bar{3}m$), has been reported to exhibit a direct band gap, albeit one where the optical transitions at the band-gap are weak.[7] Perturbation however of the hexagonal lattice by the introduction of Si or intriguingly through volume expansion of the hexagonal lattice can substantially enhance the optical transitions at the band-gap.[8,9] The Si incorporation route, eased by the fact that Si and Ge readily react with each other[10] and that both elements exhibit a hexagonal modification,[11,12] has recently been implemented.[13,14] The intriguing volume expansion route to high light emission efficiency in group IVA systems, with all its promise, is more challenging. Volume expansion has accordingly been considered near impossible to realize experimentally.[9] There could however potentially be a route to this. It would require incorporation of a larger element, Sn. But this has obstacles. Sn namely, unlike Si, does not react with Ge at ambient pressure[10] and also unlike Si,[11] cannot be recovered as a hexagonal modification under any conditions.[15] High pressures and temperatures are however a formidable vehicle in transforming phase relations (Figure 1), concomitantly creating routes to reactivity between elements leading to new structures.[22] But this nonetheless requires being able to stabilize new GeSn alloys specifically with hexagonal symmetry in a binary phase space where only cubic and tetragonal modifications are established to be stable.[23-25]

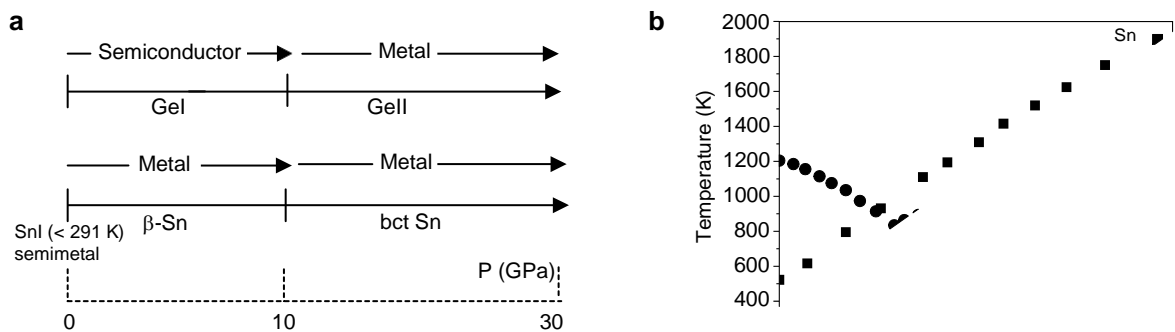

**Figure 1.** Phase relations, structural and electronic characteristics of Ge and Sn as a function of pressure and temperature. a) GeI (cubic diamond structure, $Fd\bar{3}m$) irreversibly transforms at room temperature to tetragonal GeII (β-Sn structure, $I4_1/amd$) above about 10 GPa.[16] Tetragonal Sn II (β-Sn structure, $I4_1/amd$) reversibly transforms to tetragonal Sn (bct, $I4/mmm$) above about 10 GPa.[15, 16] Sn II also transforms to Sn I (cubic diamond structure, $Fd\bar{3}m$) below about 291 K at 1 atm and below about 0.4 GPa at 291 K.[17] b) Pressure dependency of the melting points of Ge and Sn.[18-21]

Our bulk syntheses at 9 and 10 GPa and up to 1500 K employed large volume press methods. The high pressure and temperature synthetic methods were coupled with 4th generation synchrotron angle dispersive X-ray diffraction (ADX)[26] and precession electron diffraction (PED) for structural analysis, and fluorescence and electron microscopy (EM) for chemical and morphological analysis. The combination of these characterization methods is formidable for analysis of extreme conditions recovered reaction products.[5,14,27] They combine namely, high angular resolution bulk powder structural analysis (ADX), high spatial resolution single crystal analysis with kinematic intensities (PED), on-site synchrotron chemical fingerprinting using a fluorescence detector and high spatial resolution chemical and morphological mapping (EM). Experimental method details are presented in the Supporting Information.

We present below X-ray diffraction patterns from bulk Ge-Sn samples extracted from pellets recovered from large volume press experiments followed by an example of a PED pattern from a recovered crystal from a large volume press experiment. This is followed by examples of chemical analysis spectra from the recovered products. The X-ray diffraction patterns in Figure 2 reveal that hexagonal $P6_3/mmc$ Ge-Sn solid solutions can indeed be prepared using high pressures and temperatures. Even more versatile, it is possible to prepare different polytypes, namely 2H (Figure 2a), 4H (Figure 2b) as well as 6H (Figure 2c) polytypes. Figure 3 shows a PED diffraction pattern (Figure 3a) of a 2H polytype together with its simulated pattern (Figure 3b) revealing an excellent match with the experimental pattern and Figure 4 shows chemical analyses from the recovered products.

Based on this first report of synthesis of Ge-Sn hexagonal $P6_3/mmc$ alloys and previous work on this binary system, hexagonal Ge-Sn alloys can be recovered from at least as low as 9 GPa for Sn compositions up to about 16 at%. Tetragonal ($P4_32_12$) Ge-Sn alloys with up to 10 at% Sn have also been recovered from synthesis experiments at least down to 10 GPa [23]. Tetragonal ($I4_1/amd$) Ge-Sn alloys with up to about 10 at% Ge have been recovered from synthesis experiments at least down to 10 GPa and cubic ($Fd\bar{3}m$) GeSn alloys with concentrations up to about 32 at% Sn can also be recovered from high pressure and temperature synthesis experiments down to at least 7 GPa [25,29].

Based on these results we can provide an explanation for this a priori unanticipated binary new hexagonal Ge-Sn landscape. Ge is retained as a cubic diamond ($Fd\bar{3}m$) phase up to about 10 GPa upon compression (Figure 1).[16] Above this pressure it transforms to a tetragonal ($I4_1/amd$) phase.[16] This transformation is not reversible because it is kinetically hindered. Under less hydrostatic conditions the Ge $I4_1/amd$ phase is recovered with a tetragonal ($P4_32_12$) symmetry and under more hydrostatic conditions this phase is recovered with cubic ($Ia\bar{3}$) symmetry fleetingly, which then transforms to a hexagonal ($P6_3/mmc$) one. [12, 16] Sn also exhibits the cubic diamond ($Fd\bar{3}m$) phase below room temperature but transforms to the tetragonal ($I4_1/amd$) modification by heating to room temperature or by applying less than 1 GPa pressure below room temperature (Figure 1).[16] The ambient Sn ($I4_1/amd$) modification transforms to another tetragonal ($I4/mmm$) modification above 10 GPa and this transformation is reversible regardless of hydrostaticity (Figure 1). [15]

The degree of kinetic hindrance is linked to the degree of directional (non-metallic, covalent) bonding. The more directional the bonding the more kinetically hindered a phase transition can be. Cubic diamond ($Fd\bar{3}m$) Ge is a semiconductor and is comprised exclusively of directional bonding.[30-31] Also Ge's nominally metallic ($I4_1/amd$) modification largely loses its non-directionally bonded metallic character (its directional bonding is enhanced) on decompression below 10 GPa[32] causing ultimately kinetically hindered transitions to modifications with $P4_32_12$ or $Ia\bar{3}$ space groups.[16] Sn on the other hand is purely metallic in both its tetragonal modifications ($I4_1/amd$; $I4/mmm$)[30] and its cubic diamond $Fd\bar{3}m$ modification, is a semimetal.[33]

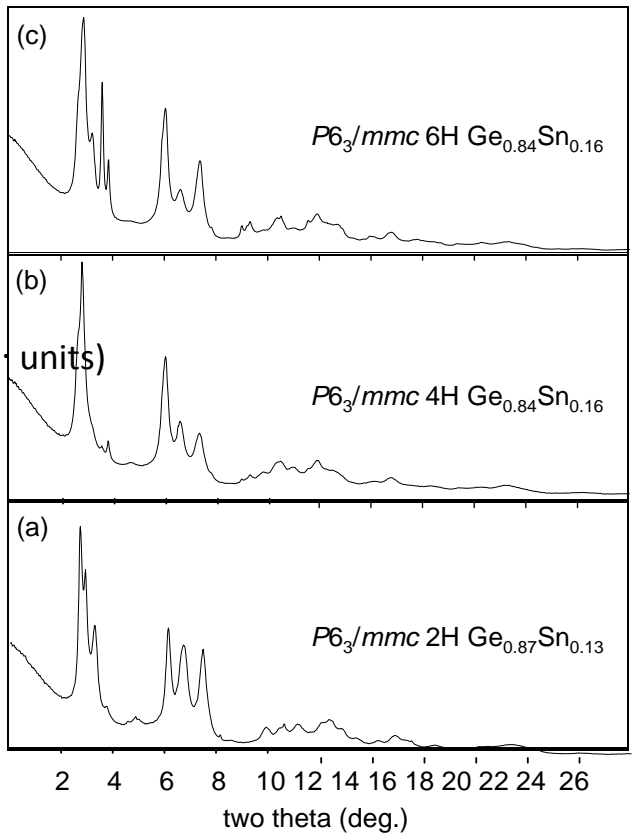

**Figure 2.** X-ray diffraction patterns of Ge-Sn solid solutions of three different hexagonal polytypes recovered at ambient conditions from high pressures and temperatures. a) X-ray diffraction pattern of a bulk $Ge_{0.87}Sn_{0.13}$ composition with $P6_3/mmc$ 2H space group recovered from 10 GPa after melting at 1500 K and annealing at 770 K [a = 4.037 Å (1), c = 6.693 Å (1); Rietveld refinement in Figure S1 in the Supporting Information].[28] b) X-ray diffraction pattern of a bulk $Ge_{0.84}Sn_{0.16}$ composition with $P6_3/mmc$ 4H space group recovered from 9 GPa after melting at 1500 K and annealing at 1000 K [a = 4.093 Å (1), c = 13.397 Å (4); Rietveld refinement in Figure S2 in the Supporting Information].[28] c) X-ray diffraction pattern of a bulk $Ge_{0.84}Sn_{0.16}$ composition with $P6_3/mmc$ 6H space group recovered from 9 GPa after melting at 1500 K and annealing at 1000 K [a = 4.092 Å (1), c = 20.081 Å (12); Rietveld refinement in Figure S3 in the Supporting Information].[28]

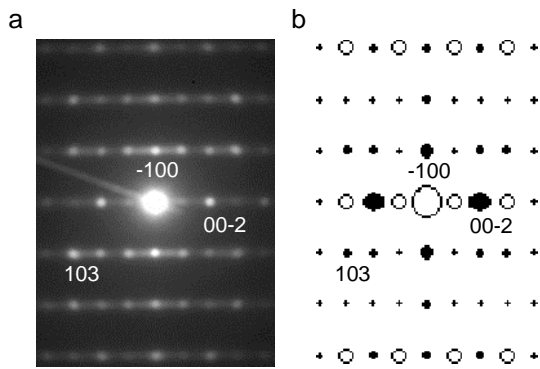

**Figure 3.** Experimental PED and simulated zone-axis diffraction patterns of a Ge-Sn crystal from multianvil synthesis with the $P6_3/mmc$ space group of 2H polytype. a) Experimental and b) simulated zone-axis diffraction pattern of the [010] zone axis.

The back-transformation of tetragonal $I4_1/amd$ Ge-Sn solid solutions to cubic diamond $Fd\bar{3}m$ solid solutions will hence be facilitated by an increased Sn presence. The minimum Sn amount required is likely more than the lower bound of 21 at% Sn we have measured in recovered Ge-Sn cubic diamond solid solutions when these are formed via back-transformation from the tetragonal ones. This is because the formation of these cubic diamond solid solutions is accompanied by ex-solution of either a pure Sn or a Sn-rich $I4_1/amd$ phase.[25] The mostly nanocrystalline nature of the recovered $Fd\bar{3}m$ solid solutions is likely due to the combined influence of the large volume change of the tetragonal to cubic transition and the ex-solution. For less than at least 21 at% Sn incorporation in $I4_1/amd$ Ge-Sn solid solutions, back-transformation from $I4_1/amd$ to $Fd\bar{3}m$ solid solutions is kinetically hindered with respect to back-transformation to either $P4_32_12$ or $Ia\bar{3}$ solid solutions which can form within this compositional regime, the latter transforming further to the current Ge-Sn $P6_3/mmc$ solid solutions.

For pure Ge, the $Ia\bar{3}$ and concomitant $P6_3/mmc$ phase is principally recovered from $I4_1/amd$ Ge in the more hydrostatic environment[34] provided using a diamond anvil cell because this high pressure method allows us to embed the sample in a soft pressure medium.[35] Conversely in a large volume press pure $I4_1/amd$ Ge is principally recovered in the tetragonal $P4_32_12$ phase because the profile conditions on decompression are typically less hydrostatic[34] than those for a sample embedded in a soft pressure medium in a diamond cell. The binary Ge-Sn system however may afford greater flexibility in this respect because Sn is more compressible than Ge.[15,36] This means that depending on the exact pressure profile conditions either tetragonal $P4_32_12$ or $Ia\bar{3}$ and concomitant $P6_3/mmc$ Ge-Sn solid solutions can be recovered in a large volume press.

The drive for new materials, especially group IVA alloys to advance information processing, clean energy and light emission applications is substantial.[13,25,37-40] The establishment hence here of a new group IVA alloy landscape with hexagonal Ge-Sn has wide ranging importance. Synthetically using Sn for preparing hexagonal group IVA alloys was unforeseen and thus also opens up new possibilities for materials development. Here, because tetragonal $I4_1/amd$ Sn is stable down to ambient pressure,[15] this could also allow preparation of $I4_1/amd$ GeSn and concomitant recovery of bulk Ge-Sn $P6_3/mmc$ solid solutions from a wider range of lower pressures at judiciously employed synthesis temperatures. Already for bulk Ge-Sn $Fd\bar{3}m$ solid solutions we have observed this recovering $Fd\bar{3}m$ solid solutions including Sn compositions between 0 and 20 at%. Moreover with Sn incorporation tunability of properties is multifaceted. Compositionally, lattice parameter variation and metallic – semiconducting solution, both affect the size of the direct band-gap.[41] Crystallographically, different polytypes can be prepared where not only the band-gap but also the structural stability can be tuned as a function of stacking sequence, with stability being likely greatest for the 6H polytype.[42] Additionally the synthetic methods employed here allow the formation of bulk and free of substrate attachment, GeSn products, which is important towards developing optoelectronic devices with integration of components on the same chip.[43] This ability to now make hexagonal GeSn polytypes, and a range of them (2H, 4H, 6H), along with cubic diamond 3C,[25] can further allow us to produce seamless free-standing heterostructures containing selected polytypes which are also of growing interest in optimally tailoring optoelectronic characteristics.[44] Additionally our ability to tailor crystal symmetry formation as a function of Sn composition provides enhanced capability to engineer phase stability.

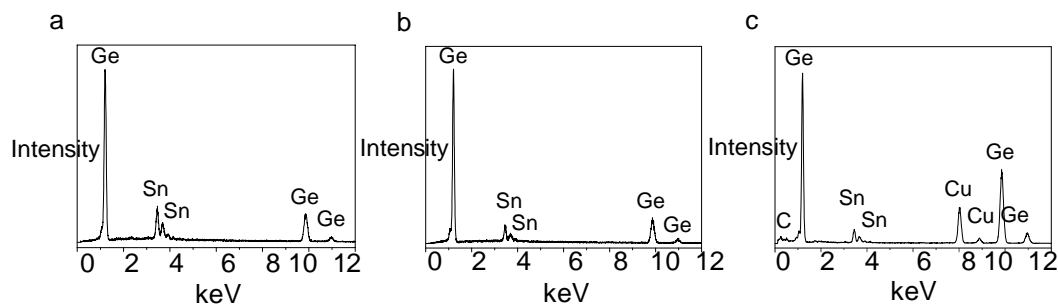

**Figure 4.** (a) Examples of SEM/EDX and TEM/EDX chemical analyses from Ge-Sn with $P6_3/mmc$ space group after recovery from high pressure and temperature syntheses. Semi-quantitative energy dispersive X-ray analysis from polished pellets and an individual crystal reveal a) $Ge_{0.84}Sn_{0.16}$, b) $Ge_{0.87}Sn_{0.13}$ and c) $Ge_{0.93}Sn_{0.07}$ stoichiometries. The Cu and C peaks originate respectively from the copper grid and the carbon film of the TEM sample holder.

## Supporting information

Detailed experimental description of syntheses, characterization methodologies, Rietveld refinements and associated data are included in the supporting information. The authors have also cited additional references within the supporting information.[45-68]

### Conflict of Interest

The authors declare no conflict of interest

### Data Availability Statement

The data that support the findings of this study are available in the supplementary material of this article.

## Acknowledgements

The TEM facility in Lille (France) is supported by the Conseil Regional du Nord-Pas de Calais and the European Regional Development Fund (ERDF). We thank Professor Jean-Paul Morniroli for his guidance and support. We gratefully acknowledge grants CH6365 and MA712 for experiments at ESRF (European Radiation Synchrotron Facility) as well as the EU "Research Infrastructures: Transnational Access" Programme (Bayerisches Geoinstitut). We further warmly thank Ahmed Addad for his work and consultation on chemical microanalysis using transmission electron microscopy, Nicola Cayzer for consultation on scanning electron microscopy and Michael Hall for demanding solids processing. We also thank Sean Tan, Andrew McGaff and Niall Russell for contributing to component micromanufacture, processing and characterization. We also thank the workshop at GFZ for manufacturing of precision holders for diffraction experiments.

## Keywords

**Entry for the Table of Contents**

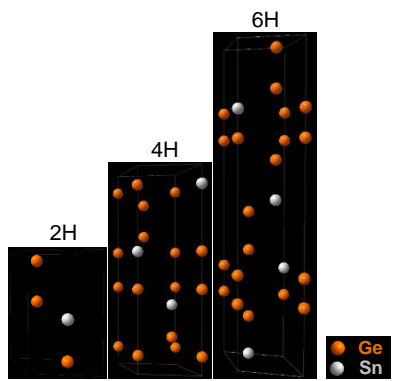

Ge-Sn is intensely investigated to overcome cubic Si's optoelectronic inadequacies. The hexagonal symmetry is of great interest in this regard. Ge and Sn however do not react in the bulk at ambient pressure and a hexagonal binary form did not exist. We remove these obstacles, preparing not just one, but a series of hexagonal GeSn alloys and stable at ambient pressure, with broader synthetic, crystal-chemical and technological implications.